   \documentstyle[aps,epsf,preprint,prl]{revtex}

\begin{document}
   \bibliographystyle{aip}
   % \documentstyle[epsf,twocolumn,prb,aps]{revtex}
   % \documentstyle[twocolumn,prb,aps,epsf]{revtex}
   % \begin{document}
   % \twocolumn[\hsize\textwidth\columnwidth\hsize\csname
   % @twocolumnfalse\endcsname % \draft \preprint{}
   \title{Aharonov--Bohm spectral features and coherence lengths in
   carbon nanotubes}
   \author{S. Roche${\ }^{\dagger}$
   G. Dresselhaus${\ }^{*}$, M. S. Dresselhaus${\ }^{+}$, and
   R. Saito${\ }^{\ddagger}$} 
   \address{ ${\ }
   ^{\dagger}$Departamento de Fisica Teorica, Universidad de
   Valladolid, E-47011 Valladolid Spain. \\ 
   ${\ }^{*}$Francis
   Bitter Magnet Laboratory, Massachusetts Institute of
   Technology, Cambridge, Massachusetts 02139\\ 
   ${\ }
   ^{+}$Department of Physics and Department of Electrical
   Engineering and Computer Science, \\Massachusetts Institute
   of Technology, Cambridge, Massachusetts 02139\\ 
   ${\ }
   ^{\ddagger}$ Department of Electronic Engineering,
   University of Electro-communications,\\ 1-5-1 Chofugaoka
   Chofu, Tokyo 182-8585, Japan}
   %\date{\today} 
   %\vspace{20pt}
   \maketitle 
   \begin{abstract}
   The electronic properties of
   carbon nanotubes are investigated in the presence of
   disorder and a magnetic field parallel or perpendicular to
   the nanotube axis. In the parallel field geometry, the
   $\phi_{0}(=hc/e)$-periodic metal-insulator transition (MIT)
   induced in metallic or semiconducting nanotubes is shown to
   be related to a chirality--dependent shifting of the energy
   of the van Hove singularities (VHSs). The effect of disorder
   on this magnetic field-related mechanism is considered with
   a discussion of mean free paths, localization lengths and
   magnetic dephasing rate in the context of recent
   experiments. 
   \end{abstract} 
   %end of \twocolumn 

   The discovery of multiwall carbon nanotubes (CNs) by
   Iijima\cite{CN-Ijima}, has triggered a huge amount of
   activity from basic research to applied
   technologies.\cite{CN-basis} Indeed, CNs have unique
   physical properties, from their light weight and record-high
   elastic modulus, to their geometry-dependent electronic
   states. Lately, there has been increasing industrial
   interest in CN properties for their applicability to flat
   displays, fiber reinforcement technologies, carbon-based
   nanotips\cite{nanotip}, and future CN-based molecular
   electronic devices.\cite{CNIC} CNs consist of coaxially
   rolled graphene sheets determined by only two integers
   $(n,m)$, and depending on the choice of chirality, metallic
   or semiconducting behavior is exhibited for systems with
   typical radii of $0.5$ to $20$\,nm and lengths of several
   $\mu$m. The study of the influence of a magnetic field and
   topological defects\cite{Charlier} or chemical (e.g.,
   substitutional)\cite{Louie} disorder is a current subject of
   concern, since disorder can strongly affect the generic
   properties of CN-based devices, such as field effect
   transistors.\cite{CN-Samsung,CN-STM,CN-FET}
   %%% The paragraph should be new here.

   Conductivity measurements have been performed on bundles of
   single-walled carbon nanotubes using scanning tunneling
   microscopy (STM).\cite{CN-RTT} By moving the tip
   along the length of the nanotubes, sharp deviations in the
   $I-V$ characteristics could be observed and related to
   theoretically predicted electronic
   properties.\cite{CN-basis,CN-WF,Rubio} In particular, the
   partition of the spectrum into a complex van Hove
   singularity (VHS) pattern is a remarkable feature. At
   energies where VHSs occur, the band velocities tend to zero,
   which is a mani\-festation of confined states, generally
   seen in 1D or 2D systems, and is due to the special symmetries
   of a given periodic structure. It is thus expected that the
   presence of VHSs (distribution and number) also affect the
   physical properties of CNs. Recently it has been shown that from
   VHS-patterns, one could distinguish between different
   nanotube chiralities,\cite{Saito-VHS} which is an
   interesting approach for obtaining a more precise knowledge
   of the effect of geometry on the physical properties of CNs,
   such as transport. 

   In this context, experiments with a
   magnetic field perpendicular\cite{langer96,naud99} or
   parallel to the nanotube axis\cite{NT-AB} have been
   performed, and in the latter case, $\phi_{0}/2$-periodic (where
   $\phi_{0}$ is a flux quantum)
   Aharonov--Bohm oscillations of the magneto-conductance have
   been found,\cite{NT-AB} suggesting rather short electronic
   coherence lengths. Fujiwara et al. \cite{AB-jap} also
   observed a surprising $\phi_{0}/3$-oscillation in the 
   magnetoconductance and related this periodic behavior to a superposition of phase
   shifts due to spectral effects, assuming three internal tubes
   with different chiralities. In the following, we propose to
   clarify the effect of a magnetic field on the
   density of states (DOS) with respect to the field orientation
   relative to the nanotube axis, and the influence that
   disorder, as featured by random site energies, may have on
   these magnetic field effects. We further derive some
   criteria for estimating the mean free path and localization
   length qualitatively in this general context.

   \section{Spectral properties of metallic and semiconducting
   CNs}
   Without a magnetic field and disorder, the electronic
   properties of nanotubes are known to be dependent on their
   chiral vector $\vec{\cal C}_{h}=(n,m)$, expressed in
   unit-vectors of the hexagonal lattice by 
   $|{\vec{C}_{h}}|=\sqrt{3} a_{\rm C-C}\sqrt{n^2+m^2+nm}$ ($a_{\rm
   C-C}=1.42$\,\AA). From a tight-binding description of the
   graphite $\pi$ bands, with only first-neighbor C-C
   interactions, the dispersion relations can be obtained by
   diagonalization of the Hamiltonian (with periodic boundary
   conditions), and for instance for the case of armchair
   $(n,n)$ nanotubes, we can write\cite{CN-basis} %
   \begin{equation} 
   \begin{array}{lll}
   \varepsilon_{\pm}(k_{\perp},k_{\parallel}) & = &
   {\displaystyle \pm \gamma_{0} \left\{ 1\pm 4\cos
   \frac{k_{\perp}a}{2} \cos\frac{\sqrt{3}k_{\parallel}a}{2}
   \right. } \\ & & {\displaystyle \hspace{2cm} \left.
   +4\cos^{2} \frac{k_{\perp}a}{2} \right\}^{1/2} } \\
   \end{array} 
   \end{equation}
   where $\gamma_{0}$ is the energy overlap
   integral between carbon atoms, $a = \sqrt{3}a_{\rm C-C}$, is
   the graphite lattice constant, and $k_{\perp}$ is the
   wavevector perpendicular to the nanotube axis $k_{\perp}={2\pi
   q}/{(\sqrt{3} n a)}$ where $q=1,\ldots,2n$, giving the
   quantized values of the wavevector in the $\vec{\cal C}_{h}$
   direction, whereas the wavevector $k_{\parallel}$ parallel
   to the nanotube axis is associated with the specification of the
   1D Brillouin zone $-\pi/ \sqrt{3} < k_{\parallel}a < \pi/
   \sqrt{3}$ which defines the 1D-band dispersion. By
   developing dispersion relations around the $\vec{K}$-points
   (i.e., close to the Fermi energy), with small $\vec{\delta
   k}=\vec{k}-\vec{K}=({2\pi}/{|\vec{\cal C}_{h}|})
   (q-{\nu}/{3})\vec{k_{\perp}}+\delta
   k_{\parallel}\vec{k_{\parallel}}$ (with $\vec{T}\cdot
   \vec{k_{\parallel}} =2\pi$ and $\vec{T}\cdot\vec{k_{\perp}}
   =0$, where $\vec{T}$ is the smallest translation vector
   along the tube axis), one finds\cite{CN-basis} 
   \begin{equation} 
   {\varepsilon}_{\pm}(\delta k)=\pm
   \frac{\sqrt{3}\gamma_{0}}{2} \sqrt{ \left\{ \frac{2\pi
   a}{|\vec{\cal C}_{h}|} \left( q-\frac{\nu}{3} \right)
   \right\}^{2} + \left| \delta k_{\parallel} \right|^{2} }
   \end{equation} % 
   where the integer $\nu$ is related to $m$
   and $n$ by $n-m=3p+\nu$ where $p=0, 1, 2,\ldots$ and $\nu = 0, \pm 1$. Two bands may
   then cross at the Fermi level according to the value of
   $\nu$. If $\nu=\pm1$ (i.e., $n-m=3p\pm 1$), one gets
   $\Delta_g={\varepsilon}_{+}(\delta k)-{\varepsilon}_{-}(\delta
   k)= 2\pi a_{\rm C-C}\gamma_{0}/|\vec{\cal C}_{h}|$ which
   defines the gap at the Fermi energy of a semiconducting
   nanotube. For $\nu=0$ (i.e., $m-n=3p$), the system is
   metallic (in the sense $\Delta_g=0$). Typically, the gap
   energy $\Delta_g$ is, respectively, in the range $1.65 < \Delta_g <
   0.27$\,eV for
   nanotube diameters $d_t$ in the range $0.5 <d_t <3\,$nm. Predicted
   theoretically,\cite{CN-basis,Ando-EB} these results have
   been confirmed experimentally by scanning tunneling spectroscopy (STM)
   measurements.\cite{CN-STM,Olk,kim99,odom98}

   \section{Magnetic field induced MIT, splitting and shifting
   of VHS{s}} 

   To investigate Aharonov--Bohm phenomena, we
   start from the Hamiltonian ${\cal H}_{{\bf k}{\bf k'}}$ for
   electrons moving on a nanotube under the influence of a
   magnetic field \cite{CN-basis}: 
   \begin{equation}
   \begin{array}{rl} 
   {\cal H}_{{\bf k}{\bf k'}}= &
   {\displaystyle \frac{1}{N}} {\displaystyle \sum_{{\bf
   R},{\bf R'}} } e^{-i ({\bf k}.{\bf R}-{\bf k'}.{\bf R'}) -
   {\displaystyle \frac{ie}{\hbar}}\Delta \varphi_{R,R'}}\\
   \times & \left\langle \psi ({\bf r}-{\bf R}) \left|
   {\displaystyle\frac{{\bf p}^{2}}{2m}} + V \right| \,
   \psi({\bf r}-{\bf R'})\, \right\rangle \end{array}
   \end{equation} 
   where the phase factors are given by
   \begin{equation}
   \Delta\varphi_{{\bf R},{\bf R'}}=
   {\displaystyle \int_{{\bf R}}^{{\bf R'}}} {\cal A}(\xi)d\xi ,
   \end{equation}
   and $|\psi({\bf r}-{\bf R'})\rangle$ is the
   localized atomic orbital, and ${\bf p}$ and $V$ are,
   respectively, the momentum and disorder operators. As shown
   hereafter, different physics is found according to the
   orientation of the magnetic field with respect to the
   nanotube axis. In the former case, the vector potential is
   simply expressed as ${\bf \cal A}= ({\phi}/{|{\cal
   C}_{h}|},0)$ in the two-dimensional $\vec{\cal
   C}_{h}/|\vec{\cal C}_{h}|$, $\vec{T}/|\vec{T}|$ coordinate system, and the phase
   factors become $\Delta \varphi_{{\bf R},{\bf
   R'}}=i({\cal X}-{\cal X'})\phi/{|{\cal C}_{h}|}$
   for ${\bf R} = ({\cal X},{\cal Y})$. This yields new
   magnetic field-dependent dispersion relations $\varepsilon
   (\delta k,{\phi}/{\phi_{0}})$. Close to the Fermi energy, this
   energy dispersion relation is affected according to
   $ k_{\perp}\rightarrow k_{\perp}+2\pi\phi/(\phi_{0}|{\cal
   C}_{h}|)$ which leads to a $\phi_{0}$-periodic
   variation\cite{Ajiki-1} of the energy gap $\Delta_g$. Such
   patterns are reported in Fig.\,1 (top) where the total
   density of states (TDOS) at the Fermi level is plotted as a
   function of magnetic field strength $\phi/\phi_{0}$
   threading the nanotube for the metallic (9,0) and the
   semiconducting (10,0) nanotubes. Note that a finite DOS is
   found in Fig.\,1, for semiconducting and metallic nanotubes as a
   function of magnetic field,
   since we consider that the
   Green's function has a finite imaginary part, that we use to
   calculate the DOS by a recursion method \cite{SR-RS}.

   %\begin{figure}[htbp]
   % \epsfxsize=8cm
   % \epsfile{file=/home/guest/roche/PRB-ABsp-Fig1.eps,width=8cm}
   %\epsfxsize 12.0cm
   %\centerline{\epsffile{/usr2/saito/roche/PRB-ABsp-Fig1.eps}}
   %\caption[]{(a) TDOS(states/eV/1C-atom) at the Fermi level as
   %a function of magnetic field parallel to the nanotube axis in
   %units of $\phi/\phi_{0}$ for $(9,0)$ (solid) and $(10,0)$
   %(dash-dot) nanotubes, and (b)TDOS(states/eV/1C-atom) at the
   %Fermi level as a function of $\mid {\cal C}_{h}\mid /2\pi
   %\ell_m$ for a magnetic field perpendicular to the nanotube
   %axis. Here $\ell_m =\sqrt{\hbar c/eB}$ is the magnetic
   %length.} \label{fig1} 
   %\end{figure} 
   \begin{figure}[htbp]
   \epsfxsize=9cm 
   \centerline{\epsffile{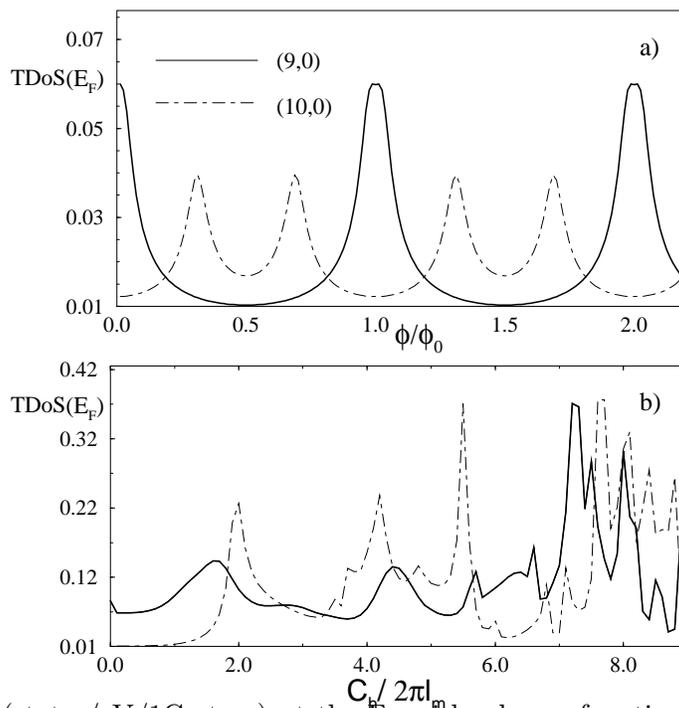}}
   \caption{(a) TDoS(states/eV/1C-atom) at the Fermi level as
   a function of magnetic field parallel to nanotube axis in
   the unit of $\phi/\phi_{0}$ for $(9,0)$ (solid) and
   $(10,0)$ (dash-dot) nanotubes, and (b)
   DoS(states/eV/1C-atom) at the Fermi level as a function of
   $\mid {\cal C}_{h}\mid /2\pi \ell_m$ for magnetic field
   perpendicular to the nanotube 
   axis. Here $\ell_m=\sqrt{\hbar c/eB}$ is the magnetic length.} 
   \label{fig1} 
   \end{figure}
   In the semiconducting case, the oscillations in the DOS
   correspond to the following variations of the gap
   widths \cite{Ajiki-1}:
   \begin{equation} 
   \label{eqn5a}
   \Delta_{g}=\left\{
   \begin{array}{lll}
   \Delta_{0} \left|1-3{\displaystyle \frac{\phi}{\phi_{0}}}\right| & \hbox{if} & 0\leq \phi 
   \leq
   {\displaystyle \frac{\phi_{0}}{2}}\vspace{1mm} \\
   \Delta_{0} \left|2-3{\displaystyle \frac{\phi}{\phi_{0}}}\right| & \hbox{if} & 
   {\displaystyle \frac{\phi_{0}}{2}}\leq \phi \leq 
   \phi_{0}
   \end{array}
   \right. 
   \end{equation} 
   where $\Delta_{0}=2\pi a_{\rm C-C}\gamma_{0}/{|\vec{\cal
   C}_{h}|}$ is a characteristic energy associated with
   the nanotube. It turns out that at $\phi$ values of
   $\phi_{0}/3$ and $2/3\phi_{0}$, in accordance with the
   values of $\nu=\pm 1$, there is a local gap-closing in
   the vicinity of either the $K$ or $K'$
   points in the Brillouin zone. This can be seen simply
   by considering the coefficients of the general
   wavefunction in the vicinity of the $K$ and
   $K'$ points, which can be written as
   $\Psi_{\vec{K}+\delta {\vec{k}}_{K} }
   (\vec{r}+{\vec{\cal C}}_{h})$. Since periodic boundary
   conditions apply in the $\vec{{\cal C}}_{h}$ direction,
   one can write $\hbox{exp} [i(\vec{K}+\delta
   {\vec{k}}_{K})\cdot{\vec{\cal C}}_{h}] =1$. For $\nu=+
   1$, we write $\delta\vec{k}_{K}=(2\pi/|{\cal
   C}_{h}|)(q-1/3)
   \vec{k_{\perp}} + \delta k_{\parallel} \vec{k_{\parallel}}$
   whereas $\delta\vec{k}_{K'}=(2\pi/|{\cal C}_{h}|)(q+1/3)
   \vec{k_{\perp}} + \delta k_{\parallel} \vec{k_{\parallel}}$.
   When $K\rightarrow K'$, then $\pm 1/3\to \mp 1/3$ in
   the above expressions, as $\nu$ goes from $+1
   \rightarrow -1$, which makes the situation between
   $K$ and $K'$ symmetrical. Similarly for metallic
   nanotubes, the gap-width $\Delta_{g}$ is expressed by
   %\begin{equation} 
   %\Delta_{g}=\left\{matrix{
   %3\Delta_{0} {\displaystyle \frac{\phi}{\phi_{0}}} & \hbox{if} & 0\leq \phi 
   %\leq {\displaystyle \frac{\phi_{0}}{2}} \cr
   %\Delta_{0} |3-3{\displaystyle \frac{\phi}{\phi_{0}}}|
   % & \hbox{if} &{\displaystyle\frac{\phi_{0}}{2}}\leq \phi \leq \phi_{0}\cr
   %}\right. 
   \begin{equation} 
   \Delta_{g}=
   \left\{
   \matrix{
   3\Delta_{0} {\displaystyle \frac{\phi}{\phi_{0}}} & \hbox{if} & 0\leq \phi \leq 
   {\displaystyle \frac{\phi_{0}}{2}} \cr
   3\Delta_{0} 
   \left| 1-{\displaystyle\frac{\phi}{\phi_{0}}} \right| 
   & \hbox{if} & 
   {\displaystyle \frac{\phi_{0}}{2}}\leq \phi \leq \phi_{0}\cr
   }
   \right. 
   \label{eqn5b}
   \end{equation} 
   Here we show that the magnetic field also affects the positions of the
   VHSs over the entire spectrum, which may be
   experimentally observed. As an illustration, the TDOS for a (10,10)
   tube without a magnetic field, shown as curve (a) in Fig.\,2, is compared to the
   magnetic field case with $\phi/\phi_{0}=0.125, 0.25, 0.375$ and $0.5$
   [shown in Fig.\,2 as 
   curves (b), (c), (d) and (e), respectively] for energies up to
   $3$\,eV in which we have taken $\gamma_{0}=2.9$\,eV, which is suitable for comparison
   with experiments.\cite{Saito-VHS,Ichida99} To understand such effects,
   one calculates the TDOS from
   \begin{equation} 
   \hbox{Tr}[\delta(E-{\cal
   H})]=\frac{2}{2\pi} \sum_{n} \int dk \delta(E-\varepsilon_{nk}),
   \label{eqn5c}
   \end{equation} 
   where the sum is taken over the $n$ energy bands
   $\varepsilon_{nk}$ and we use 
   \begin{equation}
   \left| \frac{\partial\varepsilon (\delta k,\frac{\phi}{\phi_{0}})}{
   \partial k_{{\perp}}} \right|^{-1}=\frac{2}{\sqrt{3}\gamma_{0} a}
   \frac{\left| \varepsilon(\delta k,
   \frac{\phi}{\phi_{0}})\right| }{ \sqrt{\varepsilon^{2} (\delta 
   k,\frac{\phi}{\phi_{0}})-\varepsilon^{2}_{q} }},
   \end{equation}
   where $\varepsilon_{q}$ indicates the position (energy) of the VHSs. We
   can rewrite the DOS as
   \begin{equation}
   \rho(E)={\displaystyle
   \frac{2a}{\pi \gamma_{0}\left| \vec{\cal C}_{h}\right|}}
   \,{\displaystyle \sum_{q=1}^{2n}}
   \,\delta_{q}(E,\varepsilon_{q}),
   \end{equation}
   where $\delta_{q}(E,\varepsilon_{q}) = |E| /
   \sqrt{E^{2} - \varepsilon^{2}_{q}}$ for
   $|E|>|\varepsilon_{q}|$ and zero
   otherwise.\cite{White2}
   \begin{figure}[htbp] 
   \epsfxsize=9cm 
   \centerline{\epsffile{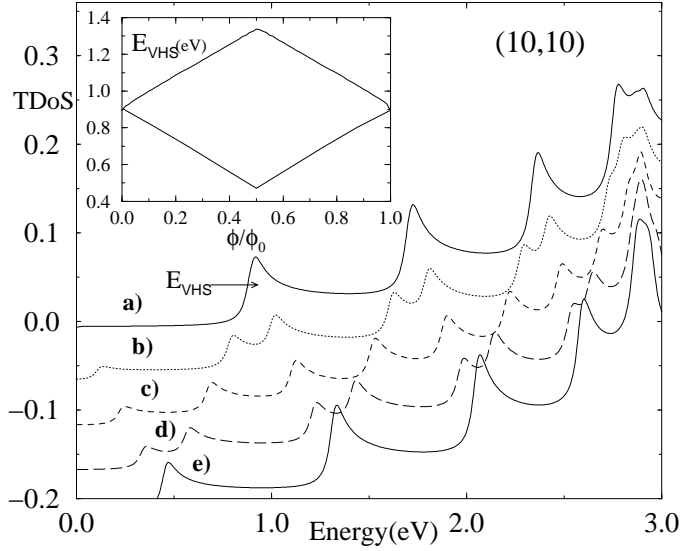}} 
   \vspace{5pt}
   \caption{TDOS [states/eV/1C-atom] of a metallic armchair (10,10) 
   nanotube as a function of Fermi energy
   for $\phi/\phi_{0} $ $=$ $(a) 0.0$, $(b) 0.125$, $(c) 0.25$, $(d)
   0.375,$ and $ (e) 1/2$. Note that curves $b$, $c$, $d$ and $e$
   have been downshifted for the sake of clarity. The
   inset shows the evolution of the first VHS position (energy in
   units of eV)
   over the complete oscillation cycle, initially located
   at $E= 0.896$\,eV for $\phi/\phi_{0}=0.$\\  
   } 
   \label{fig2} 
   \end{figure}
   %\begin{figure}[htbp] 
   %\epsfxsize=8cm 
   %\epsfxsize 12.0cm
   %\centerline{\epsffile{/usr2/saito/roche/PRB-ABsp-Fig2.eps}}
   %\epsfile{file=/home/guest/roche/Europhys-Lett/PRB-ABsp-Fig2.eps,width=8cm} 
   %\vspace{5pt}
   %\caption{TDOS [states/eV/1C-atom] of a metallic armchair (10,10) 
   %nanotube as a function of Fermi energy
   %for $\phi/\phi_{0} $ $=$ $(a) 0.0$, $(b) 0.125$, $(c) 0.25$, $(d)
   %0.375,$ and $ (e) 1/2$. Note that curves $b$, $c$, $d$ and $e$
   %have been downshifted for the sake of clarity. The
   %inset shows the evolution of the first VHS position (energy in
   %units of eV)
   %over the complete oscillation cycle, initially located
   %at $E= 0.896$\,eV for $\phi/\phi_{0}=0.$.\\
   %****Something is wrong with the figure. The $E_{vhs}$ in the inset is above 0.9
   %for $\phi/\phi_0 =0$ contrary to what is says in the caption.****
   %Label eV needed on inset.
   %}
   %\label{fig2} 
   %\end{figure}
   The $\varepsilon_{q}$ denote the energy-positions of
   the VHSs, and for armchair nanotubes $(n,n)$, one finds
   that $\varepsilon_{q}=\pm
   \gamma_{0}\sqrt{1+3\cos^{2}q\pi/n}$ and $\varepsilon_{q}=\pm 
   \gamma_{0}\sqrt{1-\cos^{2}q\pi/n}$ define the whole set of VHSs ($q=1,\ldots,5$).
   The magnetic field induces a shift of $k_{q}$ by a
   factor $2\pi\phi/(\phi_{0}|\vec{\cal C}_{h}|)$, which
   results in a new expression for the quantized values of
   the wavevector in the $\vec{\cal C}_{h}$ direction,
   which read $2\pi(q+\phi/\phi_{0})/|\vec{\cal
   C}_{h}|$. For the metallic (10,10) nanotube, an energy gap
   thus opens at the Fermi level, whose width $\Delta_{g}$
   is proportional to the magnetic strength for
   $\phi/\phi_{0}\leq 1/2$. Given that $\varepsilon_{q}$
   is a function of $k_{q}$, a shift of the energy
   positions of the VHSs follows, as well as the breaking
   of the degeneracy of a pair of $K$ and $K'$ points
   contributing to a given VHS, as explained below.
   In the (10,10) nanotube, the first five VHSs (which
   each have a degeneracy of 4) are simply given by
   $E_{\rm VHS}= \gamma_{0}\sin( \pi q/10), \
   (q=1,\ldots,5) $ which leads to $E_{\rm
   VHS}^{q=1}=0.896$~eV, $E_{\rm VHS}^{q=2}=
   1.705$~eV, $E_{\rm VHS}^{q=3}= 2.346$~eV, $E_{\rm
   VHS}^{q=4}= 2.758$~eV, and $E_{\rm VHS}^{q=5}=
   2.900$~eV. Then phase shifts induced by the magnetic
   field correspond simply to
   \begin{equation}
   E_{\rm VHS}^{q}= \gamma_{0}\sin
   \left\{\frac{\pi}{10}
   \left(q \pm \frac{ \phi}{\phi_{0}}\right)
   \right\}
   .
   \end{equation}
   This is illustrated in Fig.\,2, which gives the evolution of the TDOS
   of a (10,10) nanotube close to some VHS. 

   At low field, the degeneracy of the $K$ and $K'$ points is broken as
   illustrated in Fig.\,3, which shows the Brillouin zone for 2D graphite
   along with the equi-energy contours around $K$ and $K'$ points, so
   that in addition to the upshift of one component of the degenerate
   zero-field VHSs, the other magnetically-split component is
   correspondingly downshifted. Both VHSs move apart up to
   $\phi/\phi_{0}=1/2$, where the two originally different VHSs merge
   into one. The positions of such merged VHSs are related to a $\cos
   k_{q}a/2$ factor (that is, through a $\cos 2\pi\phi/\phi_{0}$ factor),
   so that a further increase of the magnetic field strength yields an
   opposite shifting of the VHS positions, and at $\phi/\phi_{0}=1$, the
   initial positions of the VHSs are recovered, along with their
   degeneracy (those given at $\phi/\phi_{0}=0$), and a gap closing thus
   results. At low magnetic field since $E_{\rm VHS}^{q}\sim\phi/\phi_{0}$, 
   the shift of VHSs position for a given
   magnetic field is proportional to $\phi/\phi_{0}$ for all VHSs.   
   In the inset of Fig.\,2, the positions of the VHSs originally at $E=0.896$\,eV
   (for zero field), is given as a function of the normalized magnetic
   flux. The splitting is illustrated and a strong shift of the lowest VHS, by
   $\sim 0.4$\,eV is predicted at half a quantum flux unit. Note that the
   upshifted and downshifted components are symmetrical with respect to the
   $\phi/\phi_{0}$-axis. 

   %\begin{figure}[htbp]
   % \epsfile{file=/home/guest/roche/Europhys-Lett/PRB-ABsp-Fig3.eps,width=8cm} 
   %\epsfxsize 12.0cm
   %\centerline{\epsffile{/usr2/saito/roche/PRB-ABsp-Fig3.eps}}
   %\caption{(a) Brillouin zone of 2D graphite 
   %and typical lines (dashed lines) defining the energy
   %bands near the $K$ points for an armchair nanotube. (b) Some
   %equi-energy contours around the $K$ and $K'$
   %points (circles) as well as an illustration of the magnetic field-induced VHS
   %shifting process. While one VHS is lower in
   %energy (closer to the Fermi energy at the $K$ and
   %$K'$ points), the other, initially symmetrical,
   %flows to higher energy.}
   %\label{fig4} 
   %\end{figure} 
   \begin{figure}[htbp] 
   \epsfxsize=8cm 
   \centerline{\epsffile{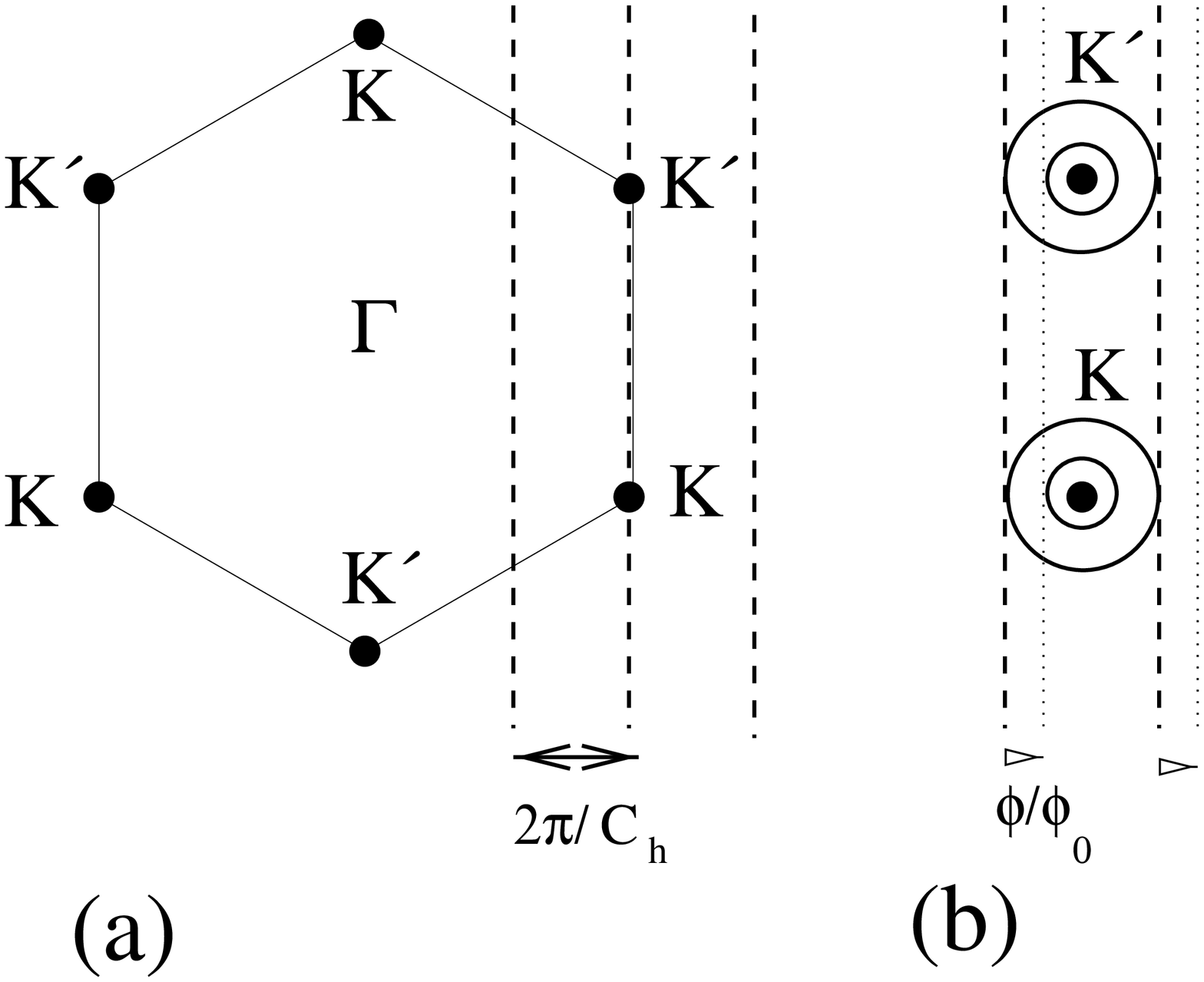}} 
   \caption{(a) Brillouin zone of 2D graphite 
   and typical lines (dashed lines) defining the energy
   bands near the $K$ points for armchair nanotube. (b) Some
   equi-energy contours around the $K$ and $K'$
   points (circles) as well as an illustration of the VHS
   shifting process. While one VHS gets closer to lower
   energy (closer to the Fermi energy at the $K$ and
   $K'$ points), the other, initially symmetrical,
   flows to higher energy.} 
   \label{fig4} 
   \end{figure} 
   \begin{figure}[htbp] 
   \epsfxsize=9.2cm 
   \centerline{\epsffile{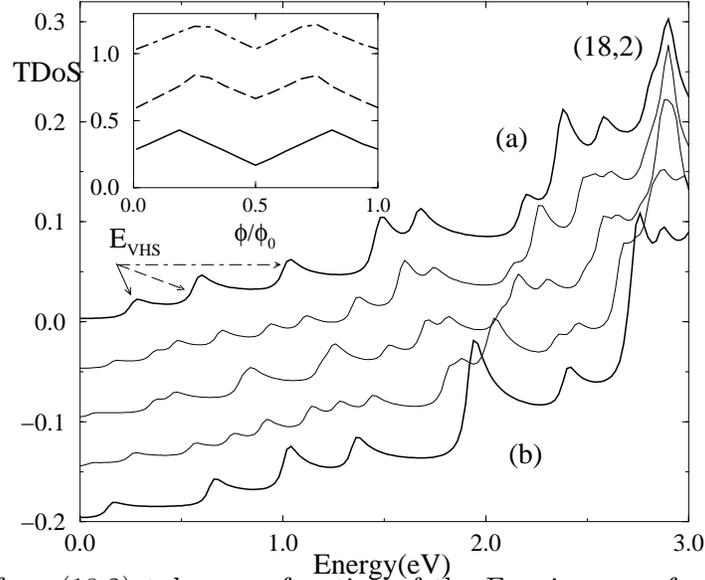}} 
   \caption{TDoS of an (18,2) tube as a function of the Fermi energy
   for several values of 
   the magnetic flux, from (a) zero flux to (b)
   $\phi_{0}/2$ flux, and intermediate cases
   corresponding to Fig.\,2. The inset shows the evolution
   of the position of three VHSs (as pointed out by
   arrows) versus normalized magnetic flux.} 
   \label{fig4} 
   \end{figure} 
   %\begin{figure}[htbp] 
   %\epsfxsize=8cm 
   % \epsfile{file=/home/guest/roche/Europhys-Lett/PRB-ABsp-Fig4.eps,width=8cm} 
   %\epsfxsize 12.0cm
   %\centerline{\epsffile{/usr2/saito/roche/PRB-ABsp-Fig4.eps}}
   %\caption{TDoS of an (18,2) tube as a function of the 
   %Fermi energy for several values of
   %the magnetic flux, from (a) zero flux to (b)
   %$\phi_{0}/2$ flux, and intermediate cases
   %corresponding to Fig.\,2. The inset shows the evolution
   %of the position (energy in units of eV) of three VHSs (as pointed out by
   %arrows) versus normalized magnetic flux.}
   %\label{fig4a} 
   %\end{figure}

   The TDOS of the semiconducting $(18,2)$ nanotube is
   also investigated and displays similar features, as
   exemplified by Fig.\,4. The inset gives the shifting of the energy
   position of the three VHSs,
   closest to the Fermi level as a function of normalized
   magnetic flux. An interesting $\phi_{0}/2$ symmetry is
   clearly seen for semiconducting tubes, as in the former
   case (Fig.\,2) for metallic nanotubes. These interesting effects
   of the magnetic field on the spectral structure has, to
   our knowledge, not been explicitly investigated
   experimentally up to now, but recent progress in the
   observation of room temperature resonant Raman spectra
   for purified single-wall nanotubes can disclose fine
   structure in the DOS, through the enhancement of the
   intensity of optical absorption at the energies
   associated with VHSs.\cite{Kataura}
   Magneto-optical techniques could then provide a possible way
   to test the above predictions. To that end, we give
   here the predicted energy position shifts for the VHSs
   closest to the Fermi energy, which are related to the
   phase shift induced by magnetic flux variations from 0
   to $\phi_{0}/2$ for nanotubes with diameters
   $d_{1}=1.8$\,nm (the largest SWNT made up to now),
   $d_{2}=6$\,nm and $d_{3}=8$\,nm (corresponding to typical
   MWNT outer diameters) and corresponding to $(n,n)$
   armchair tubes. The predicted energy position shift is given
   by
   \begin{equation} 
   \begin{array}{rl}
   \Delta E_{\rm VHS}\left(0,{\displaystyle\frac{1}{2}}\right)= &E_{\rm VHS}^{q=1}\left(
   {\displaystyle \frac{\phi_{0}}{2}}\right)-E_{\rm VHS}^{q=1}(0)\vspace{1mm}\\
   = &
   \gamma_{0}\left(\sin{\displaystyle \frac{3\pi}{2n}}-\sin{\displaystyle 
   \frac{\pi}{n}}\right) ,
   \end{array}
   \end{equation}
   which implies $\Delta E_{VHS,1} = 173$\,meV, $\Delta E_{VHS,2} =
   52.3$\,meV and $\Delta E_{VHS,3} = 39.2$\,meV, and the corresponding
   magnetic field strengths at $\phi_{0}/2$ are $B_{1}\sim 200$\,T
   $B_{2}\sim 18$\,T, $B_{3}\sim 10$\,T, which are within the scope of
   present experimental capabilities (we take
   $B=(2\pi\phi_{0})/3(na)^{2}$, and $\phi_{0}=4.1356\times
   10^{-15}$\,Tm$^{2}$). Note that the effect of the magnetic field
   should be observed for individual MWNTs, since the inter-tube coupling
   for MWNTs is believed to affect the electronic spectrum weakly.
   Whereas the splitting is universal for all tubes (chiral and achiral),
   in the most general case, the van Hove singularity shift will depend
   on the chirality, and on the applied magnetic field strength (not the
   position of the van Hove singularity). To estimate the corresponding
   shift, one may proceed in the following manner: given a chiral vector
   $\vec{\cal C}_{h}=(n,m)$, one calculates the associated vectors
   $\vec{k}_{\parallel}$ and $\vec{k}_{\perp}$ which are drawn in the Brillouin zone.
      
   This is here illustrated for the $(10,10)$ armchair for
   which
   $\vec{k}_{\perp}=(-t_{2}\vec{b}_{1}+t_{1}\vec{b}_{2})/N_{(10,10)}
   =(\vec{b}_{1}+\vec{b}_{2})/N_{(10,10)} $ with $N_{(10,10)}=40$, the
   number of hexagons within the unit cell
   of a (10,10) nanotube, and
   $\vec{b}_{i}$ the basis vector in reciprocal space (see
   Ref.\,\onlinecite{CN-basis}), where the direction $\vec{k}_{\perp}$ is
   found to be perpendicular to the $K-K'$-axis (Fig.3), and the spacing
   between lines is equidistant to a given $K$-point. For the (18,2) chiral
   nanotube with $N_{(18,2)}=364$, we find
   $\vec{k}_{\perp}=(11\vec{b}_{1}+19\vec{b}_{2})/N_{(18,2)} $ which
   defines another direction and spacing. No lines cross at $K$ and $K'$
   points, and the spacings between two consecutive lines are not
   equidistant to the $K$-location (in fact the $K$ point always appears
   to be
   one-third of the distance between the two lines near the Fermi
   energy). This affects the splitting, for instance at the field
   corresponding to half a quantum flux unit,
   the pattern for the $(18,2)$ nanotube is much less simple than that
   for the $(10,10)$ nanotube but it can
   be evaluated systematically. %%%%%%%%%%%%%%%%%%%%%%%%%%%%% 

   %%%%%%%%%%
   PERPENDICULAR CASE %%%%%%%%%%%%%%%%%%%%%%%%%%%%% 

   For a magnetic field
   perpendicular to the nanotube axis, the situation is more cumbersome,
   due to a site-position dependence of the vector potential. No apparent
   symmetries for the Aharonov--Bohm interferences on the nanotube are
   found in this case, and indeed, even if semiconducting nanotubes can become
   metallic with increasing magnetic field strength,
   non-periodic-oscillations are found, as described below. For $\vec{B}$
   normal to the tube axis, one starts from a vector potential, given in
   the two-dimensional coordinate system $(\vec{\cal C}_{h},\vec{T}$), by
   \begin{equation} 
   {\bf \cal A}= \left(0, \frac{B \left| \vec{\cal
   C}_{h} \right|}{2\pi} \sin \frac{2\pi}{\left| \vec{\cal C}_{h}
   \right|} {\cal X},0\right) . 
   \end{equation} 
   The effect of the
   magnetic field is driven by the phase factors introduced into the
   hopping integrals between two sites ${\bf R_{i}}$ and ${\bf R_{j}}$
   (with ${\bf R_{i}}= ({\cal X}_{i},{\cal Y}_{i})$, and
   the phase factor can be deduced
   from the Peierls substitution as follows, $\Delta \varphi_{{\bf
   R},{\bf R'}}$: 
   \begin{equation} \Delta \varphi_{{\bf R},{\bf R'}} =
   \left\{ \begin{array}{lr} \multicolumn{2}{l}{ {\displaystyle \left(
   \frac{ | \vec{\cal C}_{h} | }{2\pi} \right)^{2}B \frac{ \Delta {\cal
   X} }{ \Delta {\cal Y} } \left[ \cos \frac{2\pi}{| \vec{\cal C}_{h} | }
   {\cal X} \right. }} \\ \multicolumn{1}{r}{ {\displaystyle \left. -
   \cos \frac{2\pi}{ | \vec{\cal C}_{h} | } ({\cal X}+\Delta{\cal X})
   \right] } } & (\Delta {\cal X} \neq 0)\\ {\displaystyle \frac{ |
   \vec{\cal C}_{h} |}{2\pi} B\Delta {\cal Y} \sin \frac{2\pi }{|
   \vec{\cal C}_{h} |} {\cal X} } & (\Delta {\cal X} = 0)\\ 
   \end{array}
   \right. 
   \end{equation}
   where $\Delta {\cal X}= {\cal X}_{i}-{\cal X}_{j}$ and $\Delta {\cal
   Y}= {\cal Y}_{i}-{\cal Y}_{j}$\cite{CN-basis}. In Fig.\,1 (b), we show
   the total density of states (TDOS) at the Fermi level ($E_{F}=0$) as a
   function of the effective magnetic field defined by $\nu=(L/2\pi
   \ell_m)$, where $\ell_m =\sqrt{\hbar c/eB}$ is the magnetic length. At
   low fields, the TDOS of metallic-(9,0) and semiconducting-(10,0)
   nanotubes at the Fermi level increases with the magnetic field
   strength. For higher values of magnetic field, our results are in
   agreement with previous results obtained by exact
   diagonalization.\cite{CN-basis} Also Landau bands are generated for
   values for which $\nu\geq 2$. The aperiodic fluctuations of the TDOS
   are stronger at higher fields, with occasional low values of the DOS,
   reminiscent of a non-zero DOS for the semiconducting nanotubes at the
   zero-field value.
   %\begin{figure}[htb]
   %\epsfxsize=12.5cm
   %\centerline{\epsffile{/usr2/saito/roche/PRB-ABsp-Fig5.eps}}
   % \epsfile{file=/home/guest/roche/Europhys-Lett/PRB-ABsp-Fig5.eps,width=8cm}
   %\caption{TDoS of a (10,0) nanotube as a function of the Fermi energy for different
   %values of magnetic field, perpendicular to the
   %nanotube axis and with field strengths $\nu=L/(2\pi\ell_m)$}
   %\label{figo}
   %\end{figure}
   \begin{figure}[htbp]
   \epsfxsize=9cm
   \centerline{\epsffile{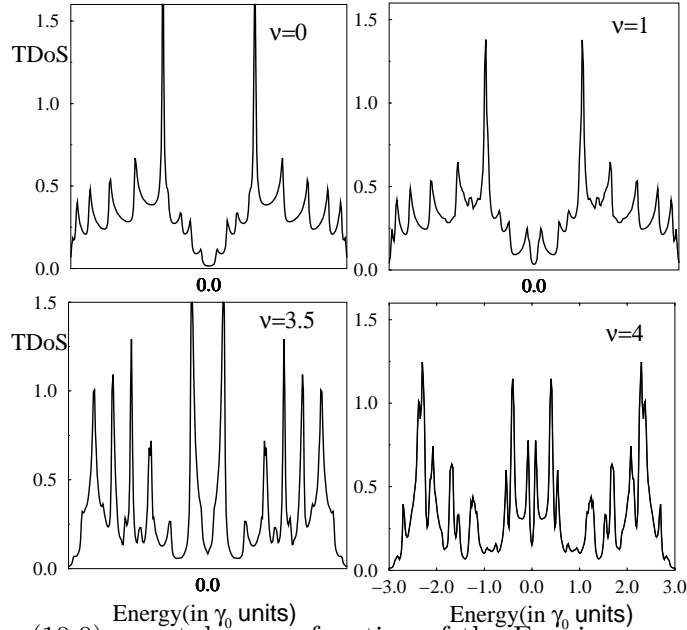}}
   \caption{TDoS of a (10,0) nanotube as a function of the Fermi energy for different
   values of magnetic field, perpendicular to the
   nanotube axis and with strength $\nu=L/(2\pi\ell_m)$.}
   \label{figo}
   \end{figure}
   In Fig.\,\ref{figo}, several values of $\nu$ are
   considered for an initially semiconducting
   nanotube. For $\nu=1$ the radius of the nanotube equals
   the magnetic length. Landau levels emerge whenever the
   magnetic length becomes smaller than the nanotube
   circumference length.\cite{Ando-EB} Comparison of the
   case of $\nu=3.5$ in Fig.\,5 with the zero-field limit
   is instructive, since the VHS partition of the spectra
   has been totally replaced by a Landau level spectrum
   (Here, one can recognize square-root singularities for
   the VHSs, and a Lorentzian-shape singularity for the
   Landau levels). 
   This transition from the VHS-pattern to the
   Landau-level pattern is, however, more unlikely to be
   observed experimentally, since its observation requires a very high
   magnetic field.

   \section{Disorder and magnetic field effects}

   The effect of Anderson-type disorder on the electronic
   properties in addition to the presence of a magnetic
   field is now addressed in order to investigate how
   disorder alters the VHS-pattern and the metal-insulator transition (MIT),
   and how disorder qualitatively modifies the
   localization properties of the nanotubes when the
   system remains metallic. We consider here only the case
   of the metallic zigzag nanotube (9,0), but similar results
   are obtained for other metallic or semiconducting
   nanotubes. In zero magnetic field, this problem of
   localization in nanotubes has recently attracted a
   great deal of
   attention.\cite{tubedis0,tubedis1,tubedis2,tubedis3}
   %For calculations, one uses recursion method described elsewhere 
   %\cite{RS-SR-PRB1}.
   \vspace{10pt}
   %$$$$$$$$$$$$$$$$$$$$$$$$$$$$$$ TRUE FIGURES
   \begin{figure}[htbp]
   \epsfxsize=9cm
   \centerline{\epsffile{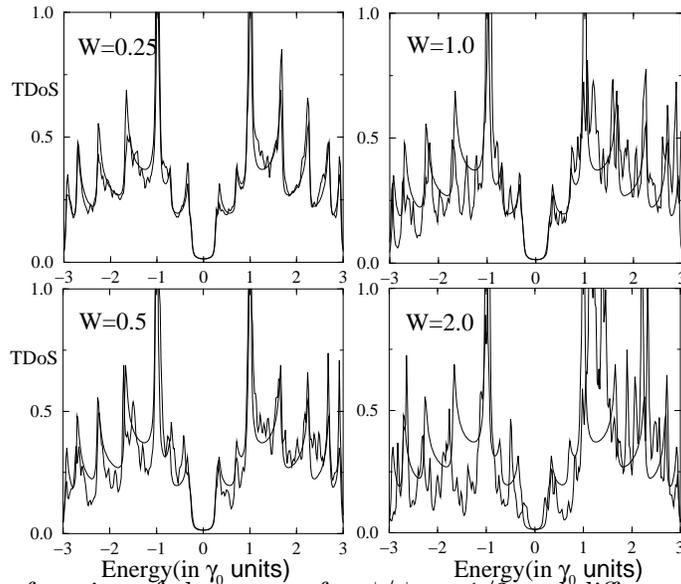}}
   \caption{TDoS as a function of the energy for
   $\phi/\phi_{0}=1/2$ and different values of the
   disorder strength $W$. Thin lines give the
   zero-disorder case for comparison. }
   \label{fig7}
   \end{figure}
   %\begin{figure}[b]
   %\epsfxsize=12cm
   %\centerline{\epsffile{/usr2/saito/roche/PRB-ABsp-Fig6.eps}}
   % \epsfile{file=/home/guest/roche/Europhys-Lett/PRB-ABsp-Fig6.eps,width=8cm}
   %\caption{TDoS as a function of the energy for $\phi/\phi_{0}=1/2$
   %and different values of the disorder strength $W$. Thin lines give
   %the zero-disorder case for comparison. } 
   %\label{fig7} 
   %\end{figure}
   The effect of randomness is considered by taking the site energies
   of the tight-binding Hamiltonian at random in the interval
   $[-W/2,W/2]$, with a uniform probability distribution. Accordingly,
   the strength of the disorder is measured by $W$. Significant
   fluctuations in the local density of states (LDOS) are found as
   shown in Fig.\,\ref{fig7} (for $\phi/\phi_{0}=1/2$). The TDoS for
   $W=0.25$ up to $W=2$ ($1/3$ of the total bandwidth) shows that
   disorder does not modify the gap at the Fermi level, even when the
   confinement effects disappear (vanishing of VHSs). Disorder
   obviously leads to a mixing of energy levels which results in a
   vanishing of the VHS at $W=1$ in our simulations. The gap is more
   resistant to disorder, even when its strength is as large as $1/3$
   the total bandwidth. Thus magnetic effects are not affected by low
   levels of disorder, which is an interesting result, since it is
   believed that many defects or sources of weak scattering should be
   present in real nanotubes. The estimation of the electron mean free path
   ($\ell_e$) in nanotubes is relevant for determining which is the
   most likely transport regime (ballistic versus diffusive), and
   further tells us whether or not the conductance should be
   quantized. While some experiments have observed conductance
   quantization,\cite{MWNT-QQ} others suggest, in contrast, rather
   short electronic coherence lengths.\cite{NT-AB} A calculation of
   $\ell_e$ for armchair nanotubes has been performed using the
   special symmetries of these systems.\cite{Todorov} Here, we show
   that a simple recourse to the relaxation time approximation (RTA)
   in the limit of weak scattering is sufficient to give a fair
   estimate of $\ell_e$ for both chiral or achiral tubes, in
   qualitative agreement with prior results,\cite{Todorov}
   demonstrated for the armchair case. Within the RTA, one can write
   $\ell_e=v_{F}\tau_{e} = \hbar v_{F}/(2\Im m\Sigma(E_{F}))$, where
   $v_{F}$ is an average of the group velocity over the Fermi surface,
   $\tau_{e}$ is the mean free time, and $\Sigma (E_{F})$ is the self energy
   due to scattering events. The calculation of the electronic
   velocity is performed by a linearization of the dispersion
   relations in the vicinity of the Fermi level. For all metallic
   nanotubes, one finds to the lowest order of approximation that
   $\sqrt{\langle v_{k}\rangle^{2}}=v_{F}=\sqrt{3} a\gamma_{0}/2
   \hbar$ which leads typically to $v_F = 8.5\times
   10^{5}$\,ms$^{-1}$. On the other hand, in the vicinity of the Fermi
   level, the density of states is given by 
   \begin{equation}
   \begin{array}{rcl} 
   \rho(E) & = & \hbox{Tr}[\delta(E-{\cal H})]\\ &
   = & {\displaystyle \frac{2}{\tilde{\ell}} \sum_{n} \int dk
   \delta(k-k_{n})\times \left| \frac{\partial
   \varepsilon_{nk}}{\partial k} \right|^{-1} } \\ &=&{\displaystyle
   \frac{2}{\pi \gamma_{0} \sqrt{n^{2}+m^{2}+nm}} } 
   \label{eqr1}
   \end{array} 
   \end{equation} 
   where $\tilde{\ell}=\Delta k|\vec{\cal
   C}_{h}|/2\pi$ is the volume of $k$-space per allowed $k$ value, divided
   by the spacing between lines for these allowed
   values. Equation\,(\ref{eqr1}) gives the TDOS per unit length of
   metallic CNs along the direction of the tube axis. Application of
   the Fermi golden rule yields $\Im m \Sigma (E_{F}) \sim \langle
   \varepsilon_{i}\rangle ^{2}\langle G_{0}(i,i,E_{F})\rangle\sim \pi
   W^{2}\rho(E_{F})$, where $W$ is the disorder bandwidth and $\langle
   G_{0}(i,i,E_{F})\rangle$ is the average of the local on-site Green
   function elements. An estimate of the carrier mean free path
   (identical for metallic nanotubes with the same radius) is:
   \begin{equation} 
   \ell_e\simeq \frac{3\sqrt{3} a {\gamma_{0}}^{2}}{2
   W^{2}}\sqrt{n^{2}+m^{2}+nm}. 
   \end{equation} 
   The important result
   that $\ell_e$ is proportional to the nanotube diameter is relevant to the
   fact that the DOS at the Fermi energy decreases with increasing
   diameter. Thus, for a given disorder acting as a weak perturbation
   coupling some eigenstates to others close to Fermi surface, there
   are less available states to be scattered into, as the tube
   diameter decreases, yielding an
   enhancement of the mean free path. For instance for (9,0) and
   (10,10) CNs, with respective diameters 0.7\,nm and 1.37\,nm, the
   corresponding mean-free paths are estimated to be $0.9\mu m$ and
   $1.8\mu m$, which are much larger than the circumference length and
   are about the typical length of the systems themselves (for a reasonable disorder
   $W\sim 0.2eV$ suggested by Ref.\,\onlinecite{Todorov}). In
   experiments on SWNTs, or MWNTs in the metallic-like regime
   (i.e., with an Ohmic temperature dependence of the resistivity), the
   resistivity should scale as $\rho(d_{nt})\sim 1/d_{nt}$ for a given
   temperature, where $d_{nt}$ is the tube diameter. 
   It would be interesting to analyze the departure from
   this law as the diameter of the MWNTs increases,
   since such departures would indicate a
   participation of the inner tubes to the measured transport regime. In particular,
   some activation transport process of inner semiconducting tubes may
   contribute or not, depending on the temperature, and thus producing
   superimposed $\rho(d_{nt})\sim \exp(\alpha_{0}d_{nt})$ factors
   (where $\alpha_{0}$ is a constant). One notes that the Fermi golden rule
   does not include the quantum interferences which lead to
   localization. As nanotubes are basically 1D-systems, from their
   estimated mean free paths, one can estimate the localization
   lengths at the Fermi level following the Thouless
   argument.\cite{Thouless} In thin wires, Thouless argued that by
   writing $R\sim 2\hbar/e^{2}$ for the resistance, there should be a
   transition to a localized state and an exponential increase of the
   resistance $R$. Consequently, from the resistance
   $R=1/G=1/\sigma\times L^{2-D}$ ($G$ is the conductance, $\sigma$
   the conductivity and $D$ the dimension of the system), assuming
   that the conductivity is calculated for a free electron gas in the
   wire ($\sigma=e^{2}/\hbar\times k_{F}\ell_e/3\pi^{2}$), it is
   straightforward to deduce the localization length $\xi=(2A
   k_{F}^{2}\ell_e)/3\pi^{2}$ (we take $A=L^{D-1}$ as the cross
   section of the wire and $L=\xi$ as the wire length for which
   $R=2\hbar/e^{2}$). Rewriting the last expression as
   $\xi=(A/\lambda_{F}^{2}) \ell_e$, the localization length $\xi$ is
   shown to be related to the approximate number of independent
   electrons (with spatial extension $\sim \lambda_{F}^{2}$, where
   $\lambda_{F}$ is the Fermi wavelength) which can be accommodated
   through the cross-section of a single tube times the average mean
   free path. When confinement effects appear in the direction
   perpendicular to the cylinder axis, the number of allowed states
   reduces to the number of bands crossing at the Fermi level
   (conduction modes), since quantization prevails, and thus $\xi
   \simeq 2N_{ch}\cdot \ell_e$\cite{Efetov}, with $N_{ch}=2$, the
   number of bands crossing the Fermi level for metallic
   nanotubes. Accordingly, for the $(10,10)$ nanotube, the
   localization length is estimated to be $\xi \sim 5$\,$\mu$m so that it
   is typically larger or similar to a typical nanotube length and
   thus no special effects due to localization (insulating regime) are
   to be expected in SWNTs. With respect to the magnetic field,
   renormalization group arguments suggest that the localization
   lengths will be weakly affected by the magnetic field \cite{Lerner}
   in quasi-1D systems. However, transport properties may be affected
   by the dephasing magnetic length $\ell_m=\sqrt{\hbar/eB}$ which is
   $\ell_m=25.66$\,nm or $8$\,nm, respectively, for $B=1$Tesla or
   $B=10$Tesla. Two cases may be distinguished, whenever the mean free
   path is much larger than the nanotube circumference. Indeed,
   following Beenakker and Van Houten \cite{Benn}, and noticing that
   in the case of a nanotube, the cross section of the quantum wire
   reduces to $\lambda_{F}/2$, the magnetic phase relaxation time
   $\tau_{B}$ for weak and strong magnetic field limits can be written
   as follows (for $\lambda_{F}\ll \ell_e$):

   \begin{eqnarray}
   \tau_{B} \simeq
   \left\{
   \begin{array}{lll}
   12\left( {\displaystyle \frac{\ell_m}{\lambda_{F}} }\right)^{2}\tau_{e}
   % }
   & {\rm if} &
   {\displaystyle
   \ell_m\ll\sqrt{\frac{\lambda_{F}\ell_e}{2}}
   }\\
   {\displaystyle 
   128
   \left(\frac{\ell_m^{4}}{\lambda_{F}^{3}\ell_e}\right)\tau_{e}
   }
   & {\rm if} &
   {\displaystyle 
   \ell_m \gg \sqrt{\frac{\lambda_{F}\ell_e}{2}}
   } \\
   \end{array}
   \right.
   \end{eqnarray}
         
   So for the considered $(9,0)$ and $(10,10)$ tubes, we
   find that $\lambda_{F}\ell_e/2\sim 18.24$nm and $25.8$nm
   respectively (we take $\lambda_{F}=0.74$nm\cite{Rubio} and
   $W=0.2078$eV\cite{Todorov}). Thus for $B=1$Tesla we conclude that
   $\ell_m\simeq \sqrt{\lambda_{F}\ell_e/2}$ which is an intermediate
   situation between low and strong magnetic field, and no analytical
   expression of $\tau_{B}$ can be deduced, whereas $B=10$Tesla is
   closer to the strong magnetic field limit where magnetic phase
   relaxation can be estimated analytically by 
   \begin{equation}
   \tau_{B}={\displaystyle \frac{12\gamma_{0}\hbar}
   {(\lambda_{F}W)^{2}}}\ell_m^{2}\sqrt{n^{2}+m^{2}+nm} 
   \end{equation}
   which gives for the $(10,10)$ armchair nanotube a dephasing rate of
   $\tau_{B}\sim 2.8\times 10^{-11}s$. This means that for an $(10,10)$
   armchair tube, the electronic phase is randomized by a 10 Tesla
   magnetic field roughly every 30\,ps, which indicates that in the
   strong field limit, the dephasing rate due to a magnetic field is just
   a few times the mean free times (we find $\tau_{B}/\tau_{e}\sim 3.14$
   with the aforementioned parameters), so that $\tau_{B}$ should
   strongly contribute to damp the quantum interferences in the weak
   localization regime. Actually, both the mean free time and the
   magnetic dephasing rate scale linearly with diameter. Inelastic
   dephasing rates due to electron-phonon coupling have been recently
   evaluated by Suzuura and Ando\cite{Suzuura}, and an interesting
   chirality-dependent dominant inelastic backscattering mechanism
   (breathing, stretching versus twisting\cite{CN-basis}) was
   revealed. From their analytical estimate of the electron-phonon
   inelastic scattering rate ($\tau_{el-ph}$), in the $(10,10)$ armchair
   case, one finds at room temperature $\tau_{el-ph}\simeq 1.34\times 10^{-5}s$ which
   is much larger than former estimated coherence times (elastic mean
   free time and magnetic dephasing rate), so that one deduces that the
   main damping effect for the weak localization regime is likely to be dominated
   by the magnetic field strength.

   Finally, with regard to the experimental results\cite{NT-AB}, one
   notices that if electronic transport is conveyed only by the outer
   shell of a metallic-like nanotube, as the magnetic field tends to
   decrease the TDoS at discrete values of the magnetic field
   corresponding to $(2n+1)/2\phi_{0}$ with integer $n$, an increase of
   resistance may follow from increasing the magnetic field. However, the short electronic
   coherence lengths that are observed in a magnetic field, the negative
   magnetoresistance and the $\phi_{0}/2$ Aharonov--Bohm oscillatory
   pattern are unlikely to be fully due to spectral effects (on the density of states),
   and a calculation of diffusion coefficients is mandatory to account for
   interferences between propagating electronic pathways in the nanotube
   structure. A recent study of the Kubo formula has been performed in
   Ref.\,\onlinecite{Roche-AB} and gives a geometrical explanation for
   the enhanced backscattering in MWNTs.

   \acknowledgments
   S.R. acknowledges the European NAMITECH Network for
   financial support [ERBFMRX-CT96-0067 (DG12-MITH)]. Part
   of the work by R.S.\ is supported by a Grant-in Aid for
   Scientific Research (No.\,11165216) from the Ministry
   of Education and Science of Japan and the Japan Society
   for the Promotion of Science for the international
   collaboration. The MIT authors acknowledge support
   under NSF grants DMR\,98-04734 and INT\,98-15744.

   \vfill\eject

\noindent
   \begin{thebibliography}{99}
   \bibitem{CN-Ijima} 
   S. Iijima, Nature {\bf 354}, 56 (1991). 
   \bibitem{CN-basis} 
   R.~Saito, G.~Dresselhaus, and M.~S. Dresselhaus, {\it 
   Physical Properties of Carbon Nanotubes} (Imperial College Press, London,
   1998). 
   \bibitem{nanotip}
   Y.~Zhang et al., Science {\bf 285} (1999) 1719.
   %T.~Ichihashi, E.~Landree, F.~Nikey and S.~Iijima
   \bibitem{CNIC}
   H.~T.~Soh et al., Appl. Phys. Lett. {\bf 75} (1999) 627.
   %C.~Quate, A.~Morpurgo, C.~Marcus, J.~Kong and H.~Dai,
   \bibitem{Charlier} 
   J.~C.~Charlier, T.~W.~ Ebbesen and Ph.~Lambin, Phys. Rev. B.
   {\bf 53} (1996) 11108.
   \bibitem{Louie}
   S.~G.~Louie, S.~Froyer and M.L.~Cohen, Phys. Rev. B {\bf 26} (1982) 1738.
   \bibitem{CN-Samsung}
   Y.~Saito, S.~Uemura and K.~Hamaguchi, J. Appl. Phys. {\bf 37} (1998) L346.
   \bibitem{CN-STM}
   J.~W.~G.~Wild{\"o}er, L. C. Venema, A. G. Rinzler,
   R. E. Smalley and C. Dekker, Nature {\bf 391} (1998) 59.
   %L.~C.~Venema, A.~G.~Rinzler, R.~E.~Smalley and C.~Dekker
   \bibitem{CN-FET}
   %Single-wall and multiwall {\sc Carbon Nanotube Field Effect Transitor (CNFET)}
   R.~Martel, T.~Schmidt, H.~Shea, T.~Hertel and Ph.~Avouris, 
   Appl. Phys. Lett. {\bf 73}, 2447 (1998). 
   \bibitem{CN-RTT}
   S.~J. Tans, R.~M. Verschueren, and C.~Dekker, Nature {\bf 393}, 
   49 (1998).
   \bibitem{CN-WF}
   L.~Venema et al., Nature {\bf 396} (1999) 52.
   \bibitem{Rubio}
   A.~Rubio et al., Phys. Rev. Lett. {\bf 82}, 3520 (1999).
   %D.~Sanchez-Portal, E.~Artacho, P.~Ordejon and J.~M.~Soler 
   \bibitem{Saito-VHS} 
   R.~Saito, G.~Dresselhaus and M.~S.~Dresselhaus, 
   Phys. Rev. B {\bf 61} (2000) 2981.
   \bibitem{langer96} 
   L.~Langer et al., Phys. Rev. Lett. {\bf 76}, 479 (1996).
   %V.~Bayot, E.~Grivei, J.~P. Issi, J.~P. Heremans, 
   %C.~H . Olk, L.~Stockman, C.~{Van~H}aesendonck, and Y.~Bruynseraede, 
   \bibitem{naud99} 
   C.~Naud, G.~Faini, D.~Mailly and H.~Pascard, C. R. Acad. Sci. Paris, t. 327, 
   Serie II b (1999). 
   \bibitem{NT-AB}
   A.~Bachtold et al., Nature {\bf 397}, 673 (1999). 
   %C.~Strunk, J.~P.~Salvetat, J.~M.~Bonnard,L.~Forro, T.~Nussbaumer
   %and C.~Sch\"onenberger
   \bibitem{AB-jap}
   A.~Fujiwara et al., Phys. Rev. B. {\bf 60}, 13492 (1999).
   %K.~Tomiyama, H.~Suematsu, M.~Yumura and K.~Uchida, 
   \bibitem{Ajiki-1}
   %Electronic states of Carbon nanotubes
   H.~Ajiki and T.~Ando, J. Phys. Soc. of Japan {\bf 62}, 
   1255 (1993). J.~P. Lu, Phys. Rev. Lett. {\bf 74}, 1123 (1995). 
   \bibitem{Olk} 
   C.~H.~Olk and J.~P.~Heremans, J. Mater. Res. {\bf 9} (1994) 259.
   \bibitem{kim99} 
   P.~Kim, T.~Odom and J.~L.~Huang, Phys. Rev. Lett. {\bf 82} (1999) 1225.
   Phys. Rev. Lett. {\bf 82}, 1225 (1999). 
   \bibitem{SR-RS}
   S.~Roche and R.~Saito, Phys. Rev. B59 (1999) 5242.
   \bibitem{odom98} 
   T.~W.~Odom et al., Nature {\bf 391} (1998) 62.
   % \Name{J.P. Lu}
   % \Review{ Phys. Rev. Lett.} \Vol{74} \Year{1995} \Page{1123}.
   \bibitem{Ichida99} 
   M.~Ichida, J. Phys. Soc. Japan {\bf 68} (1999)
   3131.
   \bibitem{White2}
   J.~M.~Mintmire and C.~T.~White, Phys. Rev. Lett. {\bf 81}, 2506 (1998). 
   \bibitem{Ando-EB}
   %energy bands of carbon nanotubes in magnetic fields
   H.~Ajiki and T.~Ando, Journ. Phys. Soc. of Japan {\bf 65}, 505
   (1996).
   \bibitem{Kataura}
   H.~Kataura et al., Synth. Met. {\bf 103} (1999) 2555.
   %Y.~Kumazawa, Y.~Maniwa, I.~Umezu, S.~Suzuki, Y.~Ohtsuka 
   %and Y.~Achiba,
   \bibitem{tubedis0}
   L.~Chico, L.~X.~Benedict, S.~G.~Louie and M.~L.~Cohen, Phys. Rev. B {\bf 54}, 2600 (1996).
   \bibitem{tubedis1}
   M.~P.~Anatran and T.~R.~Govindan, Phys. Rev. B {\bf 58}, 4882 (1998).
   \bibitem{tubedis2}
   T.~Kostyrko, M.~Bartkowiak, and G.~D.~Mahan, Phys. Rev. B {\bf 60}, 10735 (1999)
   \bibitem{tubedis3}
   K.~Harigaya, Phys. Rev. B 60 1452 (1999).
   \bibitem{MWNT-QQ}
   S.~Frank, P.~Poncharal, Z.~L.~Wang and W.~A.~de Heer, Science {\bf 280} 
   (1998) 1744.
   \bibitem{Todorov}
   C.~T.~ White and T.~N.~Todorov, Nature {\bf 393} (1998) 240.
   \bibitem{Thouless}
   D.~J.~Thouless, Phys. Rev. Lett. {\bf 39} (1977) 1167.
   \bibitem{Efetov}
   K.~B.~Efetov and A.~I.~Larkin, Sov. Phys. JETP {\bf 58}, 444 (1983).
   \bibitem{Lerner}
   I.~V.~Lerner and Y.~Imry, Europhys. Lett. {\bf 29} (1995) 49.
   \bibitem{Benn}
   C.~W.~J.~Beenakker and H.~van Houten, Phys. Rev. B {\bf 38}, 3232 
   (1988).
   \bibitem{Suzuura}
   H. Suzuura and T. Ando, Mol. Cryst. Liq. Cryst. (in press). 
   \bibitem{Roche-AB}
   S.~Roche et. al., submitted.
   \end{thebibliography}
   \end{document}